\documentclass[%
amsmath,amssymb,
aps,
pra,superscriptaddress,twocolumn
]{revtex4-2}

\usepackage{amssymb}
\usepackage{graphicx}
\usepackage{dcolumn}
\usepackage{bm}
\usepackage{amsmath}
\usepackage{lipsum}
\usepackage{amsthm}
\usepackage{mathrsfs}
\usepackage{mathbbol}
\usepackage{braket}
\usepackage{physics}
\usepackage{eqparbox}
\usepackage[colorlinks, citecolor=blue, anchorcolor=blue, linkcolor=blue,urlcolor=blue]{hyperref}

\begin{document}

\title{Ground-state Properties and Bogoliubov Modes of a Harmonically Trapped One-Dimensional Quantum Droplet}
\author{Xucong Du}
\affiliation{Department of Physics and Key Laboratory of Optical Field Manipulation of Zhejiang Province, Zhejiang Sci-Tech University, Hangzhou 310018, China}
\author{Yifan Fei}
\affiliation{Department of Physics and Key Laboratory of Optical Field Manipulation of Zhejiang Province, Zhejiang Sci-Tech University, Hangzhou 310018, China}
\author{Xiao-Long Chen}
\affiliation{Department of Physics and Key Laboratory of Optical Field Manipulation of Zhejiang Province, Zhejiang Sci-Tech University, Hangzhou 310018, China}
\author{Yunbo Zhang}
\email{ybzhang@zstu.edu.cn}
\affiliation{Department of Physics and Key Laboratory of Optical Field Manipulation of Zhejiang Province, Zhejiang Sci-Tech University, Hangzhou 310018, China}

\date{\today}

\begin{abstract}
We study the stationary and excitation properties of a one-dimensional quantum droplet in the two-component Bose mixture trapped in a harmonic potential. By constructing the energy functional for the inhomogeneous mixture, we elaborate the extended the Gross-Pitaevskii equation applicable to both symmetric and asymmetric mixtures into a universal form, and the equations in two different dimensionless schemes are in a duality relation, i.e. the unique parameters left are inverse of each other. The Bogoliubov equations for the trapped droplet are obtained by linearizing the small density fluctuation around the ground state and the low-lying excitation modes are calculated numerically. It is found that the confinement trap changes easily the flat-top structure for large droplets and alters the mean square radius and the chemical potential intensively. The breathing mode of the confined droplet connects the self-bound and ideal gas limits, with the excitation in the weakly interacting Bose condensate for large particle numbers lying in between. We explicitly show how the continuum spectrum of the excitation is split into discrete modes, and finally taken over by the harmonic trap. Two critical particle numbers are identified by the minimum size of the trapped droplet and the maximum breathing mode energy, both of which are found to decrease exponentially with the trapping parameter.

\end{abstract}
\maketitle

\section{Introduction}\label{S1}
The ultradilute quantum droplet is a novel quantum state whose self-bound nature arises from the competition between two different interactions \cite{petrov2015,petrov2016}. Beyond-mean-field (BMF) interactions, also known as quantum fluctuations or Lee-Huang-Yang (LHY) corrections \cite{LHY}, play a crucial role in such a system. However, since the BMF term represents the next-order correction, it is essentially a small quantity compared to its mean-field (MF) counterpart. The interactions in a Bose gas are highly tunable, allowing the MF term to be significantly reduced or even eliminated \cite{PRL.121.173403,PRL.126.230404,PRR.3.033247}. Therefore, multicomponent or dipolar systems with bosons are promising platforms for the study of quantum droplets \cite{Barbut2019}. Experimental realizations of quantum droplets have been achieved in binary \cite{Science2018,PRL.120.135301,PRL.120.235301,PRR.1.033155} and dipolar \cite{Schmitt2016,PRL.116.215301,PRX.6.041039} Bose systems. Theoretically, the form of the BMF term in a two-component mixture depends on the dimensionality of the system. Three-dimensional (3D) BMF term is effectively repulsive, while in lower dimensions it can be either attractive or repulsive. Especially in the one-dimensional (1D) case, the system can exhibit soliton-like features as the interaction strength varies \cite{petrov2020}, and the transition from soliton to droplet behavior has been studied \cite{PRA.97.053623,PRL.120.135301}. Recent work focuses on the superfluidity and vortex states in 2D and 3D quantum droplets \cite{PRA.98.013612,PRA.98.063602}, collective excitations \cite{Huihu2020,Reimann2021}, collision dynamics \cite{PhysRevLett.122.090401,Hu2022,PhysRevResearch.3.043139}, dimensional crossover\cite{PRA.98.051603}, and the universality and metastability in the droplet system \cite{PRA.99.023618,PRL.122.193902}. 
In 1D self-bound droplet, the collective excitation spectrum has been firstly calculated by means of linearizing the eGPE around the ground state \cite{petrov2020} and the rotational properties of reveal simultaneous rigid-body and superfluid
behavior when the droplet is put in a ring-shaped confinement carrying an angular momentum \cite{PhysRevA.105.033319}. Furthermore, the quantum Monte Carlo method is applied to the 1D droplet system \cite{PhysRevLett.122.105302,PhysRevA.102.023318} and novel phases such as the pair superfluid droplets are found in the Bose-Hubbard chain of droplets when loaded into a 1D optical lattice \cite{PhysRevLett.126.023001,PhysRevResearch.2.022008}. Some very elegant review articles summarize the latest developments at the frontiers of this new state of matter \cite{Kartashov2019,Luo2020,Bottcher_2021}.

The presence of an external potential is known to transform a simple BEC system into a non-uniform one, thereby changing its properties \cite{BECandSF,Pethick}. In dipolar systems, the anisotropic nature of the interactions means that different shapes of external potentials can lead to significantly different results, such as the ground states and stability regions in the phase diagrams \cite{PRA.82.023622,PRL.98.030406,Koch2008,Lahaye2009}. Several techniques, including hydrodynamic models \cite{Dalfovo1999,PRL.77.2360,PRL.78.1838,PRA.56.R2533}, variational approach \cite{PRL.77.5320,18dy}, and Bogoliubov theory \cite{PRL.80.1134,petrov2020}, are adopted to examine how the introduction of interactions in many-body systems affects the low-lying collective excitations of trapped BECs.

The gaseous condensate is typically achieved in experiments by confining atoms in harmonic potential. Even in experiments with self-bound droplets, as shown in \cite{PRL.120.135301,PRL.120.235301,Science2018}, researchers strive to reduce the strength of the trap as much as possible to approximate the self-bound state in free space. Therefore, it is crucial to study the influence of weak external potentials on 1D self-bound droplets. 
Two recent papers have focused on the effect of the external potential on the properties of Bose mixture droplets in the (quasi-) one-dimensional case, one of which uses an extended Gross-Pitaevskii equation (eGPE) that takes into account BMF corrections to study the bistable features of symmetric trapped droplets‘ chemical potential varies with the number of particles, and validates its results by dynamical evolution \cite{PhysRevA.107.043307}. The other adopts the non-perturbative many-body Hamiltonian to investigate the correlated dynamics of collective excitations by quenching the trapping frequency \cite{PhysRevA.107.023320}. These are good attempt, however, a unified theory of excitation spectrum is still lacking for symmetric and asymmetric 1D droplets, and the stationary properties and how the collective excitation transitions from a self-bound droplet to an ideal gas with the trapping strength and particle number is still not clear. Starting with a 1D trapped droplet system, in this paper we propose two different but complementary dimensionless equations to bridge the gap between self-bound droplets and ideal gases. We focus on how the particle number and trapping parameter affect the ground state properties and the excitation modes of the system.

The remaining part of this paper is organized as follows. In next section \ref{S2}, we introduce the appropriate time-dependent extended Gross-Pitaevskii equation for this binary 1D quantum droplet in a harmonic external potential and present two equations in mutually dual dimensionless forms. In section \ref{S3}, the details of the numerical approach based on the eGPE for the ground states and the Bogoliubov theory for low-lying excitation modes are introduced. We show how the stationary properties in the ground state are affected by the trapping potential in section \ref{S4} and the excitation spectrum, especially the breathing mode of trapped droplet, is analyzed within several energy scales and the trend of breathing mode for weak and strong confinements and small and large particle numbers in section \ref{S5}. Finally, we conclude our main results in the last section \ref{S6}.

\section{Model} \label{S2}

The energy density of a 1D quantum droplet emerging in the homogeneous two-component Bose mixture is written as \cite{petrov2016,petrov2020}
\begin{align}
{\cal E}_{\text{1D}} =& \frac{(g_{11}^{1/2} n_{1} - g_{22}^{1/2} n_{2} )^2}{2} + \frac{g \delta g (g_{22}^{1/2} n_{1} + g_{11}^{1/2} n_{2} )^2}{(g_{11} + g_{22} )^2} \nonumber \\
&- \frac{2 \sqrt{m} }{3 \pi \hbar } (g_{11}n_{1} + g_{22} n_{2} )^{3/2}, \label{energydensity}
\end{align}
where $n_{1,2}$ denotes the density of the two components in different hyperfine states of the same atom of mass $m$, $g_{11,22}$ and $g_{12,21}$ represent the repulsive intra- and attractive inter-component interaction strengths with $g_{12}=g_{21}$, the parameters $g$ and $\delta g$ are related to the coupling constants in the two components as $g=\sqrt{g_{11} g_{22}}$ and $\delta g = g + g_{12}>0$, respectively.

The first two terms in Eq. (\ref{energydensity}) are the contribution of the MF theory, and the last one is the LHY correction arising from the quantum fluctuation. Under the condition of minimizing the dominant first MF term in Eq. (\ref{energydensity}), the density of these two components will follow the ratio $n_1/n_2 = \sqrt{g_{22}/g_{11}}$, which is also true in the inhomogeneous mixture for smooth enough variation of the total density. This allows us to reduce the mixture to an effective single-component system characterized by the wavefunction $\psi(x)$, which is related to that of each component as
\begin{align}
\psi_{1,2}(x) = \left( \frac12 \sqrt{\frac{g_{22,11} }{ \mathcal{G} }}\right)^{1/2}\psi(x). \label{psi-rel}
\end{align}
where we further define $\mathcal{G} = (\sqrt{g_{11}} + \sqrt{g_{22}})^2/4$. This definition assures that $|\psi(x)|^2=|\psi_1(x)|^2+|\psi_2(x)|^2$, i.e. the density of the effective single component is the sum of each component in the mixture $n(x)=n_1(x)+n_2(x)$. The energy functional for the inhomogeneous mixture is taken as 
\begin{align}
{\cal E}[\psi(x),\psi^*(x)]=&\int dx \left( \frac {\hbar^2}{2m} |\nabla \psi|^2 + V(x) |\psi|^2 \right.\notag \\
&\left. +\frac{g\delta g}{4\mathcal{G}} |\psi|^4 - \frac{2\sqrt{m}}{3\pi \hbar} g^{3/2} | \psi|^{3} \right). \label{energyfunctional}
\end{align}
Minimizing the energy functional Eq. (\ref{energyfunctional}) with respect to independent variations of $\psi(x)$ and its complex conjugate $\psi^*(x)$ subject to the condition that the total number of particles $N=\int dx |\psi(x)|^2$ be constant, we obtain the equation of motion of the system or the time-dependent extended Gross-Pitaevskii equation (eGPE) for 1D droplet in a harmonic trapping potential
\begin{align}
i\hbar \partial_t \psi =& \left[-\frac{\hbar^2}{2m} \nabla^2 + \frac12 m\omega_x^2 x^2 \right. \notag \\
& \left. +\frac{g\delta g}{2\mathcal{G}} |\psi|^2 - \frac{\sqrt{m}}{\pi \hbar} g^{3/2} | \psi| \right] \psi, \label{eGPE}
\end{align}
where $\omega_x$ is the harmonic trapping frequency. Note this equation applies equally to the binary mixture for both symmetric ($g_{11}=g_{22}$) and asymmetric ($g_{11} \ne g_{22}$) interaction strengths. 

To get a neat dimensionless form of the eGPE, we need to rescale the length, time, and wave function, respectively, as
\begin{align}
	x = x_0 x'\text{, }t=t_0 t'\text{, }\psi = \psi_0 \psi',
\end{align}
where the prime denotes the dimensionless quantities and the subscript 0 indicates the characteristic units. Note that all terms in the square bracket of (\ref{eGPE}) are energies and the energy and time units can be derived from the definition of the length unit $x_0$ using the following relationship:
\begin{align}
E_0 = \frac{\hbar^2}{mx_0^2} = \frac{\hbar}{t_0}.
\end{align}
For the purpose of making the final dimensionless equation contain as few adjustable parameters as possible, we first require that the two nonlinear terms have the same coefficients, i.e., they need to satisfy
\begin{align}
	\frac{g\delta g}{2\mathcal{G}}\psi_0^2 = \frac{\sqrt{m}}{\pi\hbar}g^{3/2}\psi_0,
\end{align}
which results in a characteristic factor for the wave function as
\begin{align}
	\psi_0 = \frac{2\mathcal{G} \sqrt{mg}}{\pi\hbar \delta g}. \label{psi0}
\end{align}
Hence the crucial step is the choice of length unit $x_0$, for which there are two different conventions to follow in the field of cold atoms and we will refer to as $\lambda$-scheme and $\beta$-scheme, respectively.
The $\lambda$-scheme is commonly used in the self-bound quantum droplet systems as suggested in \cite{18dy,petrov2020}: the length unit $x_{\lambda}$ is derived by equating the kinetic energy and the interaction energy
\begin{align}
\frac{\hbar^2}{mx_{\lambda}^2} = \frac{\sqrt{m}}{\pi\hbar}g^{3/2}\psi_0,
\end{align}
which leads to
\begin{align}
x_{\lambda} = \frac{\pi\hbar^2}{mg}\sqrt{\frac{\delta g}{2\mathcal{G}}}. \label{x-lambda}
\end{align}
Thus, a scaling factor for the number of particles is generated by the normalization condition of the wave function
\begin{align}
\int |\psi'|^2 dx' =\tilde{N},
\end{align}
where $\tilde N= N/N_{\lambda}$ is the number of particles in unit of $N_\lambda$ and
\begin{align}
N_{\lambda} =\psi_0^2  x_{\lambda} = \frac{1}{\pi}\left( \frac{2\mathcal{G}}{\delta g} \right).
\end{align}
The eGPE is then cast into the following dimensionless form
\begin{align}
	i\partial_t \psi = \left( -\frac12 \nabla^2 + \frac12 \lambda^2 x^2 + |\psi|^2 - |\psi| \right)\psi,\label{eGPE-lambda}
\end{align}
where we have omitted all the prime symbols, and
\begin{align}
	\lambda =\frac{\omega_x}{\omega_\lambda} \label{lambda}
\end{align}
is the trapping potential energy in units of the interaction energy in the self-bound droplet $E_\lambda=\hbar \omega_\lambda$ \cite{petrov2020} with
\begin{equation}
\omega_\lambda=\frac{2mg^2 \mathcal{G}}{\pi^2 \hbar^3 \delta g }.
\end{equation}
The dimensionless eGPE (\ref{eGPE-lambda}) obtained in the $\lambda$-scheme can describe the influence of the external potential on the self-bound quantum droplet. When the characteristic length of the harmonic potential is equal to that of a self-bound droplet, i.e., $\lambda = 1$, the external potential is considered significant. At this point, the characteristic units defined in $\lambda$-scheme may not be as appropriate, thus we give a second path.
\begin{figure}[t]
	\centering
	\includegraphics[width=\columnwidth]{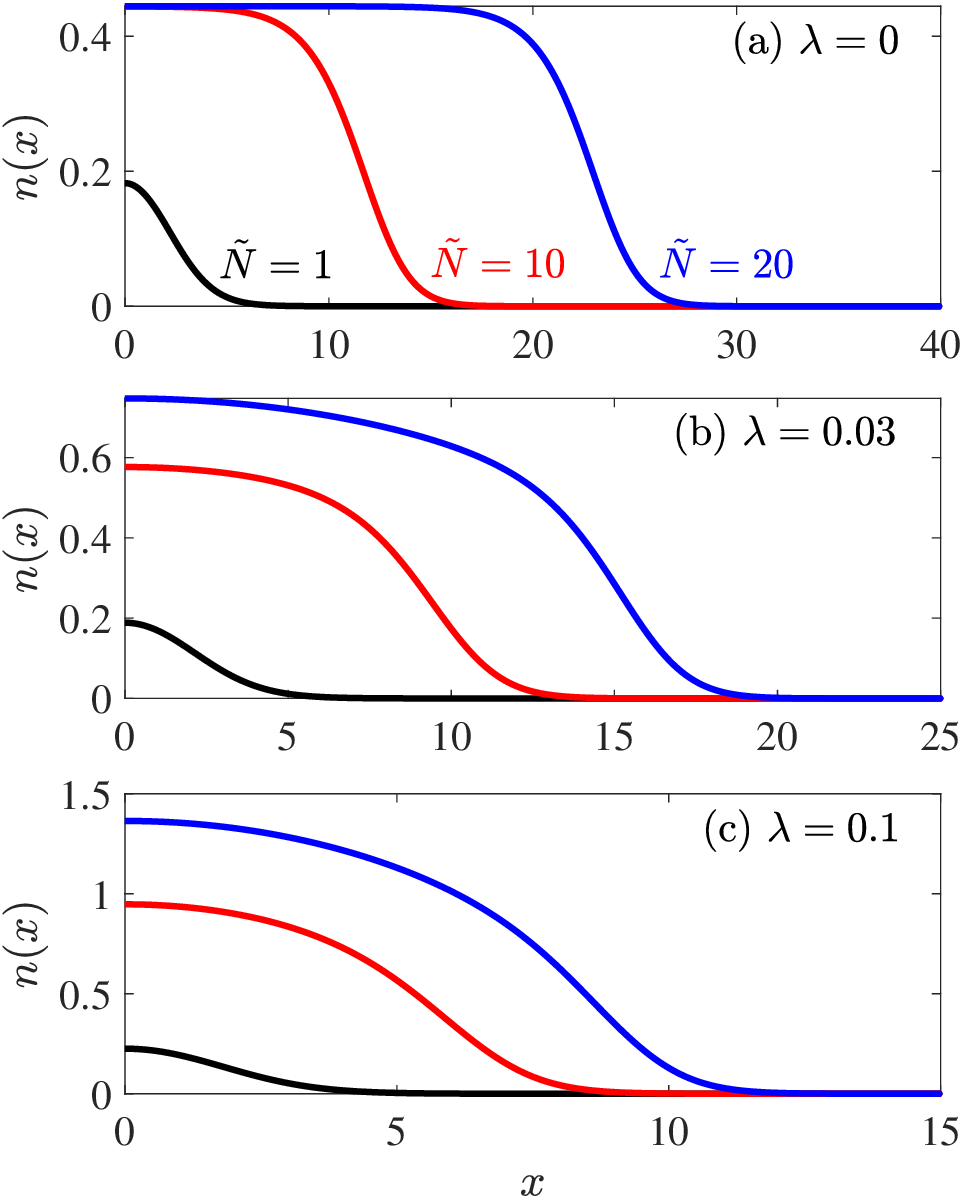}
	\caption{Density profiles of the quantum droplet for the particle number $\tilde{N}=1$ (red), $10$ (yellow), $20$ (blue), and trapping potential $\lambda = 0$ (a, top panel), $0.03$ (b, middle panel), $0.1$ (c, bottom panel), in dimensionless units defined in Eqs. (\ref{x-lambda}) and (\ref{psi0}).}
	\label{RS_Density}
\end{figure}

Once the trapping potential is strong enough, it is more reasonable to use the characteristic length of the harmonic trapping potential as the unit of length, which is actually by equating the kinetic energy and the harmonic potential
\begin{align}
\frac{\hbar^2}{mx_{\beta}^2} = \frac 12 m \omega_x^2 x^2,
\end{align}
and
\begin{align}
	x_{\beta} = \sqrt{\frac{\hbar}{m\omega_x}}. \label{x-beta}
\end{align}
Similarly, the scaling factor for the number of particles in the $\beta$-scheme is
\begin{align}
	N_{\beta} = \psi_0^2 x_{\beta}  = \frac{4\sqrt{m}g\mathcal{G}^2}{\pi^2 \hbar^{3/2}\delta g \sqrt{\omega_x}}.
\end{align}
Then the dimensionless eGPE reads (where the primes are omitted)
\begin{align}
	i\partial_t \psi = \left[ -\frac12 \nabla^2 + \frac12 x^2 + \beta (|\psi|^2 - |\psi|) \right]\psi, \label{eGPE-beta}
\end{align}
with
\begin{align}
	\beta = \frac{\omega_\lambda}{\omega_\beta}  \label{beta}
\end{align}
is the interaction energy in units of the trapping potential with $\omega_\beta=\omega_x$. The dimensionless eGPE (\ref{eGPE-beta}) obtained in the $\beta$-scheme depends solely on one dimensionless parameter $\beta$, which can equally be used to elucidate the role of the two types of interactions, i.e. MF and LHY, on the properties of a trapped droplet.

It is interesting that the dimensionless eGPE from these two schemes are in a duality relation, i.e. the parameters are inverse of each other
\begin{equation}
\beta =1/\lambda, \label{betalambda}
\end{equation}
from which the length unit and the particle number unit are related as
\begin{align}
	x_{\lambda} = \frac{x_{\beta}}{\sqrt{\beta}}\text{, }N_{\lambda} = \frac{N_{\beta}}{\sqrt{\beta}}.
\end{align}
Therefore, in order to better present the results, all physical quantities in the figures presented in this paper are dimensionless and characterized via the $\lambda$-scheme. It is worth noting that, by setting $g_{11}=g_{22}$, we naturally arrive at $\mathcal{G}=g=g_{11,22}$ and the system enters the familiar symmetric case (be careful with the definition of $n$) which are discussed in detail in Refs. \cite{18dy,petrov2016,Luo2020}.

\section{Bogoliubov theory}\label{S3}
In this section, we use the Bogoliubov method to study the low-lying collective excitations of the system. To quantitatively determine the density profile of the ground state and the collective excitations of the quantum droplet, it is necessary to solve the stationary eGPE for the ground state wave function
\begin{align}
\hat{\mathcal{H}} \psi_g = \mu \psi_g, \label{eigen-gs}
\end{align}
with $\mu$ the chemical potential in unit of $E_0$ and the operator $\hat{\mathcal{H}}$ denoted in the $\lambda$ or $\beta$-scheme as
\begin{align}
\hat{\mathcal{H}}_{\lambda} =& -\frac12\nabla^2 + \frac12 \lambda^2 x^2 + \psi_g^2 - \psi_g,\\
\hat{\mathcal{H}}_{\beta} =& -\frac12 \nabla^2 + \frac12 x^2 +\beta (\psi_g^2 - \psi_g),
\end{align}
and then solve the Bogoliubov equations for small-amplitude excitations around the condensate 
\begin{align}
\begin{bmatrix}
\hat{\mathcal{H}}-\mu + \hat{\mathcal{M}} & \hat{\mathcal{M}}\\
-\hat{\mathcal{M}} & -\hat{\mathcal{H}}+\mu - \hat{\mathcal{M}}
\end{bmatrix}
\begin{bmatrix}
	u_j\\v_j
\end{bmatrix}=\omega_j
\begin{bmatrix}
	u_j\\v_j
\end{bmatrix}, \label{BdG}
\end{align}
where we have defined the operators,
\begin{eqnarray}
\hat{\mathcal{M}}_{\lambda} &=& \psi_g^2 - \frac12 \psi_g,\\
\hat{\mathcal{M}}_{\beta} &= &\beta(\psi_g^2 - \frac12 \psi_g),
\end{eqnarray}
and used the fact that the stationary ground state $\psi_g$ is real. The Bogoliubov equations are obtained by assuming a small density fluctuation around the ground state, i.e. using the following \textit{ansatz}
\begin{align}
\psi(x,t) = \bigg\{ \psi_g(x) + \sum_j\left[ u_j(x) e^{-i\omega_j t} + v_j^*(x) e^{i\omega_j t} \right] \bigg\} e^{-i\mu t}, \label{wf_f}
\end{align}
to expand the eGPE to the first order of $u_j$ and $v_j$. We emphasize that the dimensionless Bogoliubov equations (\ref{BdG}) can be obtained either from the dimensionaless eGPEs (\ref{eGPE-lambda}) or (\ref{eGPE-beta}) by inserting back the wave function with fluctuation (\ref{wf_f}), or from the original eGPE (\ref{eGPE}) assuming a similar fluctuated wave function and performing the dimensionless after. The excitation energy (frequency) $\omega_j$ in units of $E_0$ and the amplitudes $u_j, v_j$ in units of $\psi_0$ are labeled by an integer $j$ with $j=0,1,2,\cdots$. This linearization process is a standard method for studying low-lying excitations.

In the numerical workload for obtaining the collective excitation spectrum, we first evolve the stationary eGPE in imaginary time to obtain the ground state wave function. This allows us to construct the Bogoliubov matrix, which is then diagonalized using Arnoldi's method \cite{Arnoldi1951} to obtain the eigenvalues representing the collective excitation spectrum. The implicitly restarted Arnoldi method, which is implemented in the ARPACK software package, enables finding the $M$ largest or smallest eigenvalues of the operator matrix in (\ref{BdG}), where $M$ is selected by the user. In the program we take $M = 40$, which contains 20 pairs of opposite eigenvalues, and in the end only 20 of these positive values are retained. The sine-spectral method \cite{Bao2006,Gao2020} is used to deal with the kinetic term, which is known for its higher accuracy compared to the conventional finite difference method. Specifically this method introduces the sine interpolation of a function $u(x)$ to approximate its second order derivative $\partial_{xx}u(x)$ at grid points which can be formulated as a matrix-vector multiplication.

\begin{figure}[t]
	\centering
	\includegraphics[width=\columnwidth]{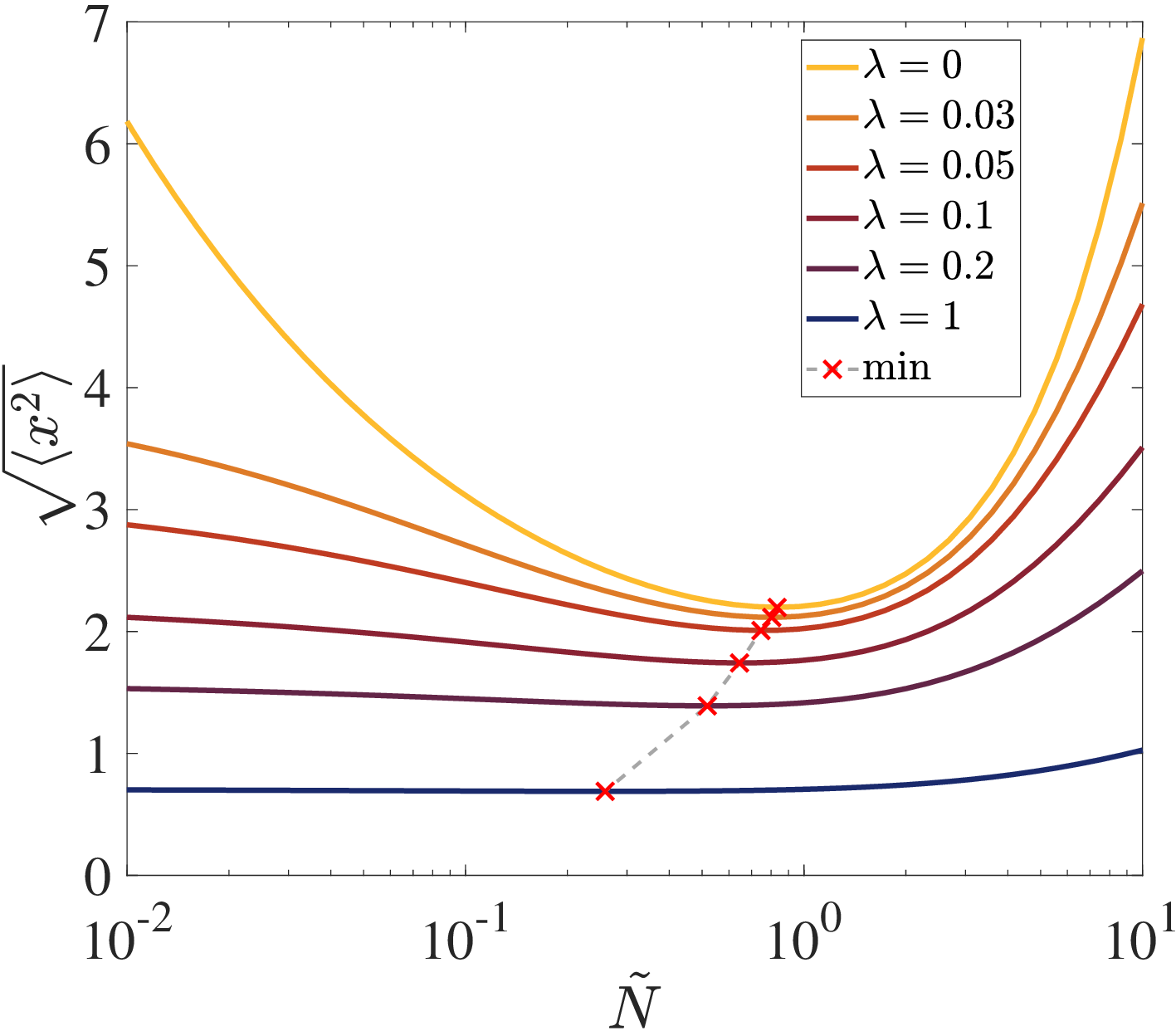}
	\caption{Mean square radius $\sqrt{\Braket{x^2}}$ of the droplet as functions of the number of particles $\tilde{N}$ for various trapping parameters $\lambda$ from the self bound case $\lambda=0$ to a strong confinement $\lambda=1$ as indicated in the legend. The minimum values marked with red crosses are critical particle numbers $\tilde N_c$ where the droplet has a minimum size.}
	\label{RS_Mr}
\end{figure}

\section{Stationary properties}\label{S4}
We first consider the effect of the applied confinement potential on the static properties of the quantum droplet from the numerical solutions of the stationary eGPE. It is already known that the density of a self-bound droplet takes a gaussian-like profile for a small number of particles, while it exhibits a flat-top structure when the number of particles is large. Here we do the numerical simulations in the $\lambda$-scheme for the reduced particle numbers in the range $\tilde{N}=0.01$ to $50$ and consider three representative values for the parameter $\lambda$: $0$, $0.03$, and $0.1$.

The density profiles of the system are shown in Fig. \ref{RS_Density}. The case $\lambda=0$ describes the self-bound droplets in free space whose density profiles undergo a smooth transition from the flat-top structure at large particle numbers $\tilde N=20$ to the Gaussian shape for a few atoms $\tilde N=1$ (Fig. \ref{RS_Density}a). The presence of an external potential introduces a slight inhomogeneity into the system and especially the central region of the droplet feels this more sensitively: more particles are pushed towards the center and the maximum density starts to increase, as shown in Fig. \ref{RS_Density}b for $\lambda=0.03$. The flat-top structures are easily destroyed and the droplet shows the bell-like profile as the strength of the external potential increases further (Fig. \ref{RS_Density}c for $\lambda=0.1$), a phenomenon that also occurs in the 3D case \cite{Huihu2020}. We observe that in the 1D case the flat-top structure is more fragile and hard to observe than its 3D counterpart, as a very weak external potential $\lambda \sim 10^{-3}$ is enough to spoil the flat-top while it is still robust for a confinement as strong as $\lambda \sim 0.03$ in 3D \cite{Huihu2020}. Note that the spatial extension in Fig. \ref{RS_Density} gradually shrink inward and the maximum value of density is squeezed to a relatively high value as the confinement becomes tighter.

\begin{figure}[t]
	\centering
	\includegraphics[width=\columnwidth]{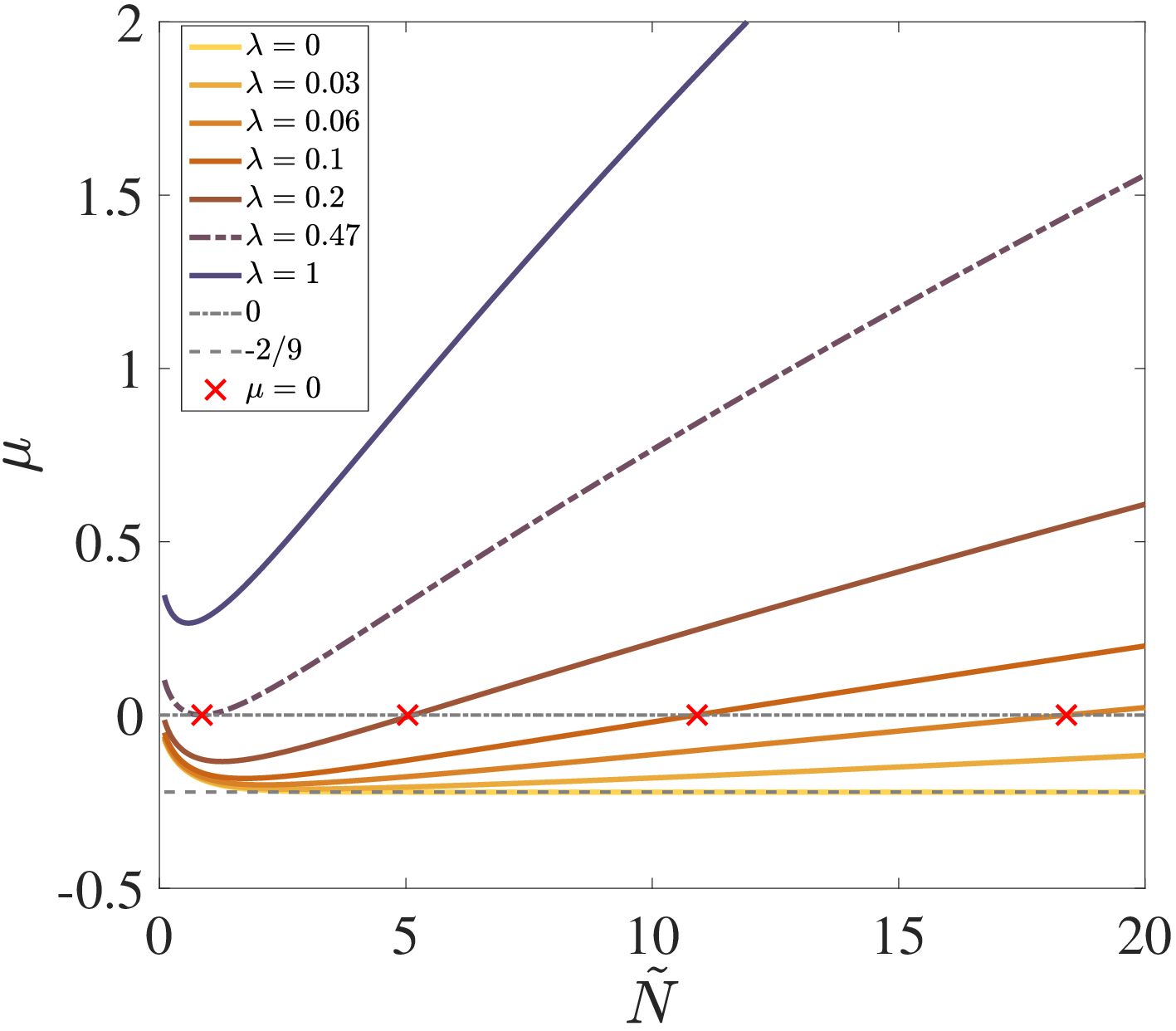}
	\caption{Chemical potential $\mu$ of the droplet as a function of $\tilde{N}$ in different traps with different parameters $\lambda$ as indicated in the legend. The dashed line represents the equilibrium value corresponding to the spatially uniform state when $\tilde N \rightarrow +\infty$, and the dot-dashed line is the negative-to-positive transition line $\mu=0$ which intersect with the chemical potential for increasing parameters $\lambda$ at fewer and fewer number of particles marked by the red corsses. The chemical potential is always positive for trapping parameter larger than $\lambda_c \sim 0.47$ (dot-dashed curve).}
	\label{RS_Mu}
\end{figure}
\begin{figure*}[t]
	\centering
	\includegraphics[width=2\columnwidth]{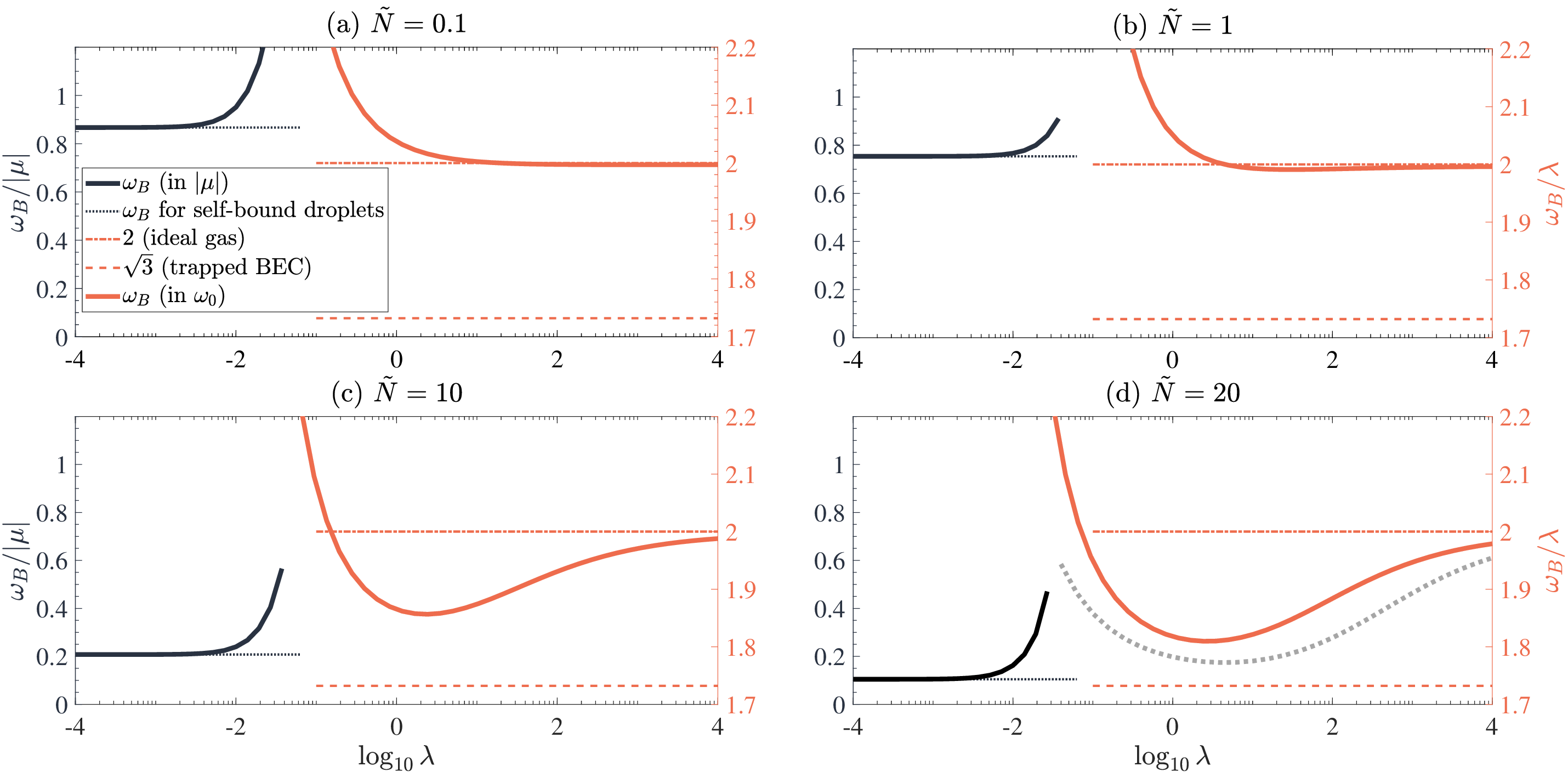}
	\caption{The ratios $\omega_B/|\mu|$ (left axis) and $\omega_B/\lambda$ (right axis) of the droplet as a function of $\lambda$ for $\tilde{N} =0.1$ (a), $1$ (b), $10$ (c), and $20$ (d). The black dashed line represents the results for the self-bound system \cite{petrov2020}, the dot-dashed lines $\omega_B=2\lambda$ and the dashed lines $\omega_B=\sqrt{3} \lambda$ are the prediction of an ideal gas and the interacting BEC system, respectively. The grey dotted curve in (d) represents the excitation energy for a very large droplet $\tilde N=50$ with a bottom very close to the value of 1D dilute bose gas.}
	\label{RS_Wb}
\end{figure*}

It is helpful to have a closer look at how the confinement would change the minimum size of the droplet. In Fig. \ref{RS_Mr} we present the numerical result for the mean-square radius $\sqrt{\Braket{x^2}}$ of the droplet with
\begin{equation}
\left< x^2 \right>=\frac{\int^{+\infty}_{-\infty} x^2 |\psi_g|^2dx}{\int^{+\infty}_{-\infty} |\psi_g|^2dx},
\end{equation}
as a function of the reduced particle number $\tilde{N}$. It can be observed that, for both the self-bound droplet and the droplet in the trapping potential, the size of the droplets exhibits a minimum value, which, marked by red crosses in Fig. \ref{RS_Mr}, in the self-bound droplet is near $\tilde{N}=1$ and moves gradually toward smaller particle numbers with increasing confinement $\lambda$. While in the self-bound case the size of a smaller droplet diverges at $\tilde N \rightarrow 0$ and in the opposite limit the size of large droplets grows linearly with $\tilde N$, we find that the confinement changes this picture, i.e. the size of a smaller droplet no longer diverges but tends to be a constant in units of $x_0$ for increasingly strong traps, and at large particle numbers the droplet size still grows linearly with $\tilde N$, but the growth rate becomes very slow. The minimal values define a critical particle number $\tilde N_c$ where the droplet has the minimum size. For the self-bound case it separates two different density profiles, i.e. the Gaussian-like profile for $\tilde N \ll \tilde N_c$ and the flap-top structure for $\tilde N \gg \tilde N_c$. But it is no longer true for the droplet in the trap as the flat-top is very fragile and easily lost for 1D trap. This critical number of particle $\tilde N_c$ depends almost linearly on the trapping parameter $\lambda$. This phenomenon is likely attributed to the dominance of the kinetic term when the number of particle is small (see \cite{18dy}). It can be explained as follows. In the self-bound system the quantum pressure generated by the kinetic term serves to balance the interaction energy. The presence of the external trapping potential then compensates for the potential energy, leading to a reduction in the required interaction energy, which also corresponds to the need for fewer particles. As a result, the minimum value of $\sqrt{\Braket{x^2}}$ gradually shifts towards lower particle numbers. This behavior is also reflected in the monopole oscillation of the collective excitation, which will be discussed in detail in the next section.

We note that the confinement will also alter the chemical potential $\mu$ greatly. For 1D self-bound droplet, the chemical potential always takes a negative value that approaches $-2/9$ for large number of particles, as shown in Fig. \ref{RS_Mu}, implying that the state is self-bound in the equilibrium in consistent with previous findings \cite{petrov2016,18dy}. In this case, the chemical potential $\mu$ can be regarded as the threshold of particle emission, making $|\mu|$ a suitable unit for excitation energy. The trapping potential contributes to the chemical potential of the 1D trapped droplet that will not only increase with the number of particles $\tilde N$ but also with the confinement parameter $\lambda$. For a weaker potential ($\lambda <0.2$), the chemical potential maintains the same downward trend as the self-bound droplets for small particle numbers, but begins to increase after a certain number of particles until its value becomes positive. For a stronger potential ($\lambda >0.5$), the chemical potential is already positive for small particle numbers and increases rapidly with $\tilde N$. We observe a critical trapping parameter $\lambda_c \sim 0.47$, above which the chemical potential is never negative. This transition of the chemical potential from negative to positive causes the ratio of excitation energy to chemical potential, e.g. $\omega_B/|\mu|$, to diverge at the transition points $\mu =0$. Marked by the red crosses in Fig. \ref{RS_Mu}, these transition points occur for smaller particle numbers for stronger trapping potentials. To avoid this divergence, in both the strong confinement limit and the large particle number limit, we plot the ratio of the breathing mode frequency $\omega_B$ and the trapping frequency $\lambda$ in the following.

\section{Breathing mode of trapped droplet}\label{S5}

\begin{figure}[t]
	\centering
	\includegraphics[width=\columnwidth]{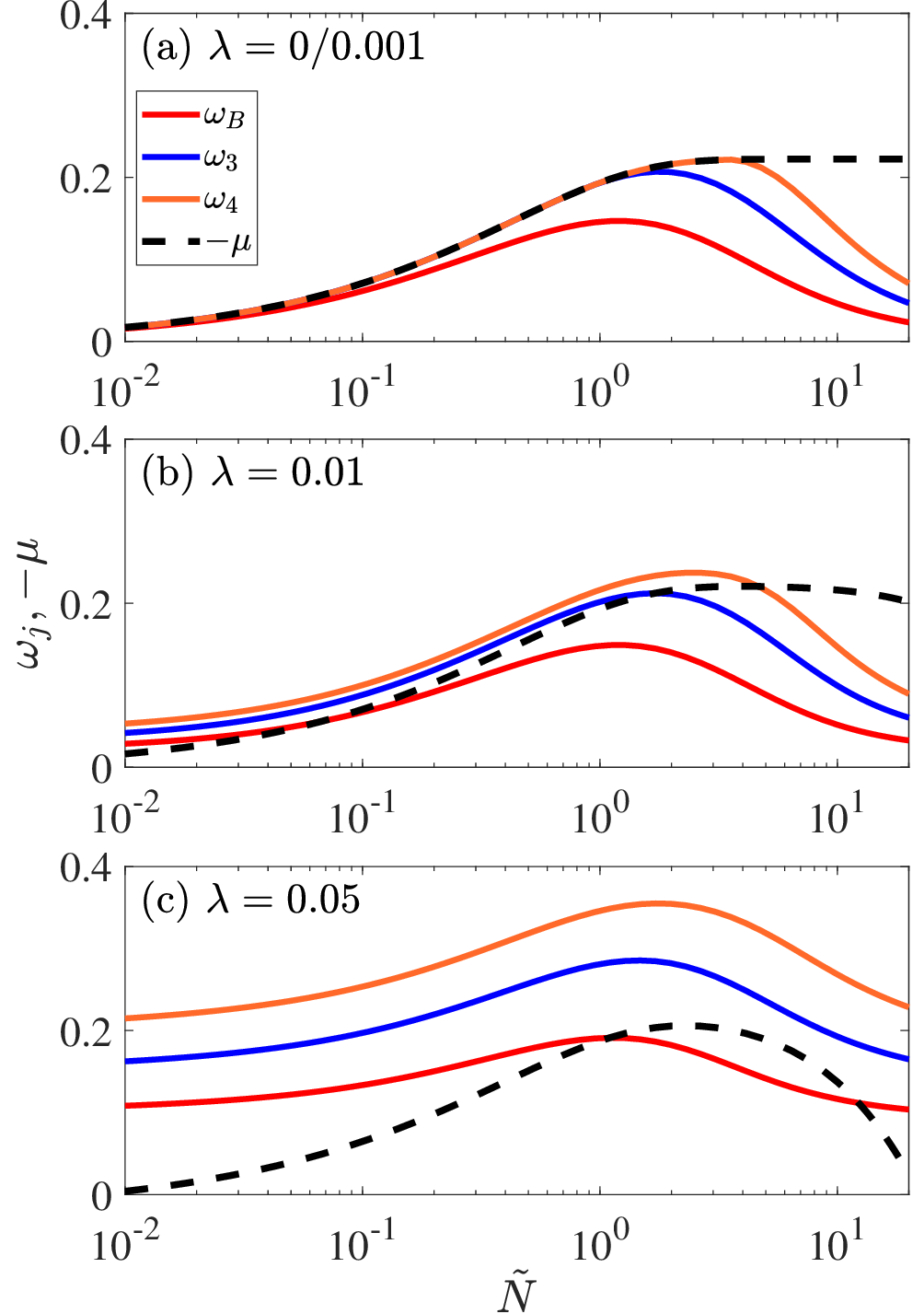}
	\caption{The first three non-trivial modes $\omega_{j=2,3,4}$ of the Bogoliubov excitation spectrum and the chemical potential (with minus sign) of the self-bound droplet and trapped droplet with different parameters $\lambda=0.001$ (a), $0.01$ (b), and $0.05$ (c) as a function of $\tilde{N}$. The dashed and solid lines are the result for the chemical potential and the excitation modes, respectively. The breathing mode $\omega_B$ is the third mode, i.e. $\omega_{j=2}$. }
	\label{RS_Wb_Mu}
\end{figure}

In the context of quantum droplets in cold atoms, the monopole oscillation, also known as the breathing mode, refers to a collective oscillation of the droplet's density profile or size, which can be calculated by means of several different methods. For example, the low-energy excitation spectrum of the system can be obtained by solving the hydrodynamic equations, where the lowest monopole mode is the breathing mode \cite{PhysRevLett.77.2360,PhysRevA.58.2385}. In the variational approach, using the time-dependent ansatz, one can define an effective potential by the Euler-Lagrange equation for the width parameter and assume that it has the form of a harmonic potential to obtain the frequency of the breathing mode \cite{18dy,otajonovPLA,OtajonovPRE,OtajonovJPB,Reimann2021}. For the time-independent situation the elegant sum-rule approach is adopted in order to evaluate the collective frequencies in the intermediate regimes where the hydrodynamic equations are not analytically soluble \cite{Huihu2020,PhysRevLett.122.070401,PhysRevA.66.043610,PhysRevA.90.013622}. In addition, the breathing mode can also be excited by introducing a small interaction perturbation \cite{Reimann2021} or by driving a single droplet out of equilibrium with an initial excitation in the norm $\tilde N$ and study the ensuing dynamics by simulating the eGPE \cite{18dy}. Here we numerically solve the Bogoliubov equation (\ref{BdG}) to extract the breathing mode of the system and check how it is affected by the confinement parameter $\lambda$. It is noteworthy that the breathing mode for the 1D droplet corresponds to the third eigenvalue of the Bogoliubov equations (\ref{BdG}), i.e. $j=2$.

The excitation energy of the breathing mode $\omega_B$ of the droplet is shown in Fig. \ref{RS_Wb} as a function of the trapping parameter $\lambda$ for several typical atomic numbers. We use two different energy units for nearly self-bound droplet on one side and very tightly confined droplet on the other side, i.e. the ratio $\omega_B/|\mu|$ and $\omega_B/\lambda$ are plotted, respectively, as explained in the last section. To illustrate more clearly the excitation energies in the two limits the logarithm coordinates are used for $\lambda$. Numerically for weak confinement $\lambda \sim 10^{-4} -10^0$ we use the $\lambda$-scheme to diagonalize the Bogoliubov equations with operators $\hat{\mathcal{H}}_{\lambda}$ and $\hat{\mathcal{M}}_{\lambda}$ and to extract the third eigenvalues, i.e. the excitation energy $\omega_B$ of the breathing mode, which are shown in black solid lines in Fig. \ref{RS_Wb}. For small or large droplets the breathing modes fall back to their self-bound values \cite{petrov2020} (the horizontal dotted lines) when the confinement is as weak as $\lambda \sim 10^{-2}$ as expected. For strong confinement $\lambda \sim 10^{0} -10^4$, on the other hand, the numerical $\beta$-scheme is preferred and it is the operators $\hat{\mathcal{H}}_{\beta}$ and $\hat{\mathcal{M}}_{\beta}$ that play the role in the diagonalization of Bogoliubov matrix and the numerical results are shown in red solid lines. We find that in very strong confinement the droplets all behave like ideal gases with the excitation energy showing a smooth convergence to the value $\omega_B/\lambda=2$ \cite{Pethick}. However, the excitation energies for larger droplets undergo a downward process with a minimum value around $\lambda \sim 1$ and eventually approach the ideal gas value from below. The downward moving of this minimum in the excitation energy is attributed to the competition among the trapping potential ($\sim N$), the mean-field ($\sim N^2$) and LHY ($\sim N^{5/2}$) interaction in the droplet and more particles would push the bottom further down to the value of weakly interacting condensate, i.e. $\omega_B/\lambda= \sqrt{3}$ \cite{PRL.77.2360}, denoted in dashed lines in Fig. \ref{RS_Wb}, and the case for a relatively large droplet of $\tilde N=50$ is additionally shown in Fig. \ref{RS_Wb}d.

\begin{figure}[t]
	\centering
	\includegraphics[width=\columnwidth]{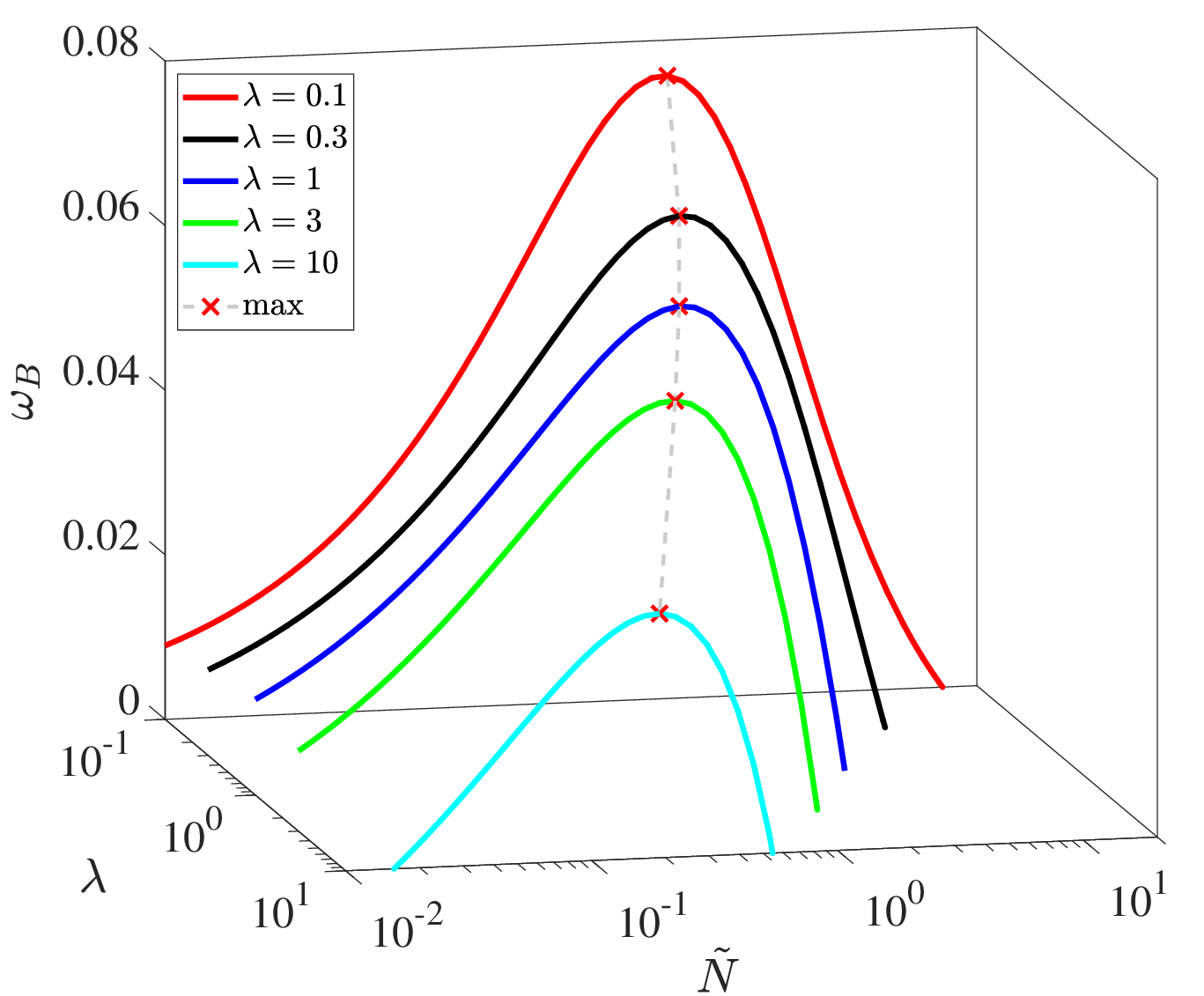}
	\caption{The dependence of the breathing mode $\omega_B$ of the trapped droplet on $\tilde{N}$ for several larger parameters $\lambda$. Just as in the self bound case, the breathing mode of trapped droplet exhibits a maximum excitation energy for each parameter $\lambda$ and determines a critical number of particles $\tilde N_c$ of $\omega_B$.}
	\label{RS_Wb_3D}
\end{figure}

The "divergence" in the excitation energy seems unreasonable, which is due to either the vanishing of the chemical potential at the transition points for $\omega_B/|\mu|$ or the disappearance of the trap for $\omega_B/\lambda$. To cure this, the dependence of the non-trivial excitation modes including the breathing mode, as well as the chemical potential, on the droplet’s size is presented  in unit of energy scale $E_\lambda$ in Fig. \ref{RS_Wb_Mu}. It is already known \cite{18dy} that for the self-bound droplet this dependence is non-monotonous with the largest excitation energy (stiffness) reached around $\tilde N_c = 1.2776$, which is very close to $\tilde N_c$ of the mean square radius when the droplet has the minimal size $\tilde N_c = 0.8330$. It is also argued that the “autocooling” mechanism, which lose atoms until all excitations are gone and droplets are generated in the true ground state for 3D droplet, is no longer applicable in 1D geometry, as the energy of breathing mode is always less than the absolute value of the chemical potential, i.e. $ \omega_B < |\mu|$. It is worth to note that there is no longer the particle emission threshold to continuum in the presence of the external trap as all excitations become bound modes with discrete frequency \cite{Huihu2020}. We find that for the 1D droplet the excitation energy is indeed lower than $-\mu$ in very weak traps as in the self-bound case \cite{18dy} - we see no difference in the excitation spectrum  for $\lambda \sim 10^{-3}$ and the self bound droplet. Note that $-\mu$ actually will approach its equilibrium value $2/9$ corresponding to the spatially uniform state when $\tilde N \rightarrow +\infty$. No intersection point is found in the lowest breathing mode and the chemical potential, while the higher modes $\omega_{j=3,4,\cdots}$ would intersect with $-\mu$ at their branching points, below which the corresponding excitations lie in the continuum spectrum, as shown in Fig. \ref{RS_Wb_Mu}a. The presence of the external trap, e.g. a very weak one with $\lambda=0.01$, immediately splits the continuum into discrete modes, which are likely to be peeled off layer by layer from the threshold $-\mu$, and allows the excitation modes of smaller droplets to increase first to exeed $-\mu$, so that we observe the intersection of the lowest breathing mode with $-\mu$, and branching points for higher modes are now intersections with $-\mu$, moving toward larger numbers of particles in Fig. \ref{RS_Wb_Mu}b. In the mean time, $-\mu$ for large $\tilde N$ is bent downward, e.g. in a trap $\lambda=0.05$ in Fig. \ref{RS_Wb_Mu}c, and two intersections appear in the competition of breathing mode and the chemical potential, but no intersections for higher modes, at the same time the discrete modes are gradually taken over by the harmonic trapping potential, i.e. for small or large $\tilde N$ the excitation spectrum tends to be equidistance with a characteristic gap proportional to $\lambda$. 
\begin{figure}[t]
	\centering
	\includegraphics[width=\columnwidth]{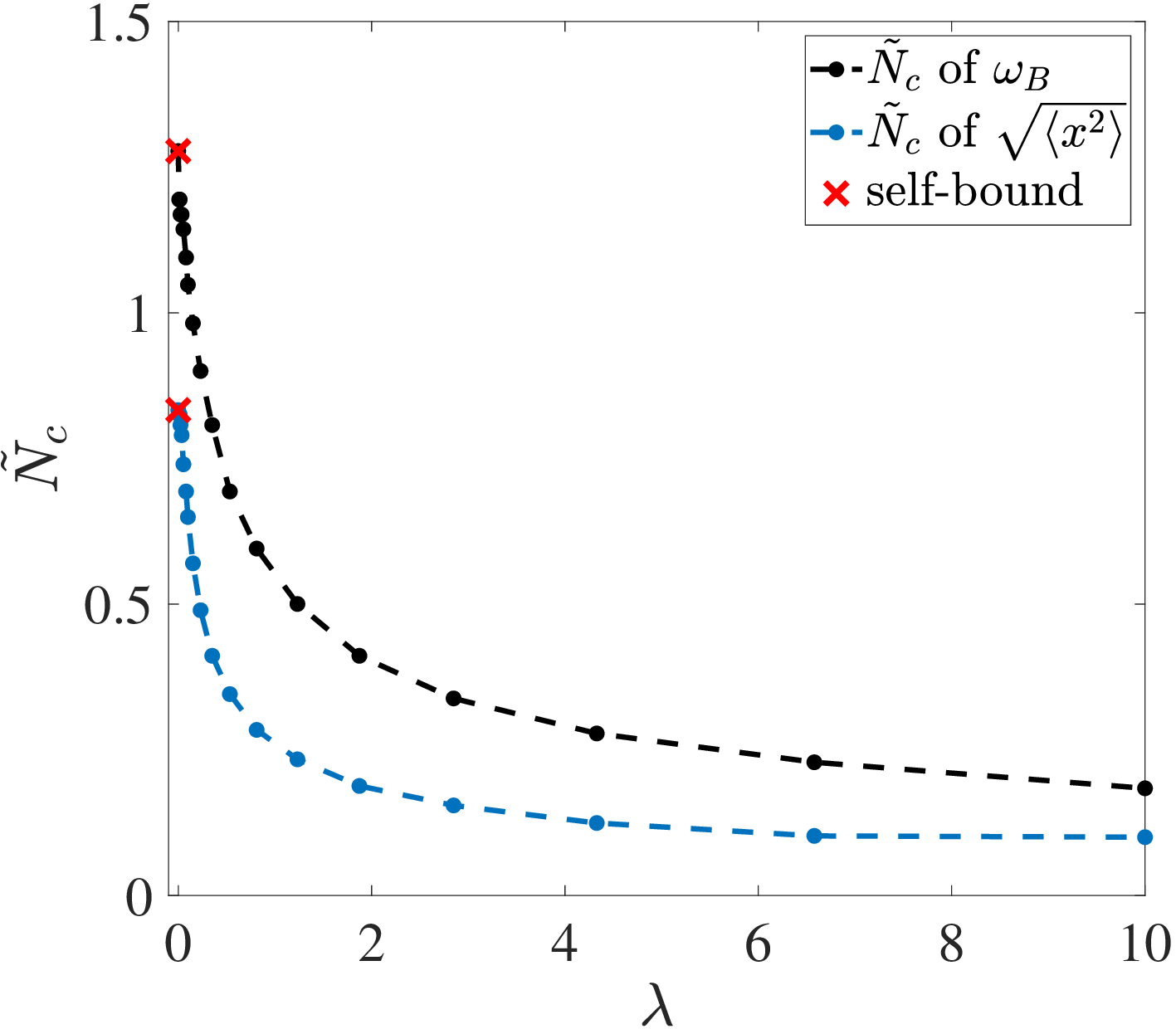}
	\caption{The critical numbers of particles $\tilde{N}_c$ determined by the minimum values of the mean square radius $\sqrt{\Braket{x^2}}$ (light blue dots) and by the maximum values of the breathing mode excitation energy $\omega_B$ (black dots) as functions of trapping parameter $\lambda$. A simple fit is shown by the dashed lines.}
	\label{RS_Nc}
\end{figure}Interestingly, for even stronger confinement, e.g. $\lambda=0.1$, the lowest excitation mode will exceed $-\mu$ completely and we further illustrate the size-dependent frequency of the breathing mode for large trapping parameters in Fig. \ref{RS_Wb_3D}, where an additional shift $2\lambda$ has been made for each mode $\omega_B$ in units of $E_\lambda$. The reason why we shift this $2\lambda$ is due to the fact that for small enough number of particles the excitation mode is more or less near the ideal gas limit $\omega_B=2\lambda$ as shown in Fig. \ref{RS_Wb}a. For each trapping parameter, there exists a peak value of number of particles, which defines another critical particle number $\tilde N_c$ where the excitation energy of the droplet $\omega_B$ has the maximum energy, marked by the red crosses in Fig. \ref{RS_Wb_3D}. Recall that the critical number of particles $\tilde N_c$ of the mean square radius in Fig. \ref{RS_Mr}, we find both critical numbers of particles $\tilde N_c$ of $\sqrt{\Braket{x^2}}$ and $\tilde N_c$ of $\omega_B$ decrease with the trapping parameter $\lambda$ exponentially, as shown in Fig. \ref{RS_Nc}, where the self-bound values of $\tilde N_c=1.2776$ and $0.8330$ are denoted by two red crosses. The discrepancy of the critical numbers in the self-bound droplet was roughly carried over into the trapped droplets, which are compared in detail in Fig. \ref{RS_Nc}. We find that clearly the existence of the minimum size of the droplet is accompanied by the appearance of the maximum excitation energy of breathing mode. It's important to note that the study of excitation modes in quantum droplets is an active and rapidly evolving field of research, and the understanding of their properties, including the breathing mode, is still developing. For example, similar minimum size for the 3D self-bound droplet occurs at $\tilde N \sim 30$, while the variational Gaussian ansatz for the breathing mode gives an estimate of $\tilde N \sim 80$ and the numerical result from eGPE is instead $\tilde N \sim 10^3$ \cite{Huihu2020}. The specific details of the droplet's composition, interparticle interactions, and external conditions can significantly influence the relationship between droplet size and the excitation energy of the breathing mode in cold atom systems.


Finally let us check the reliability of the two numerical schemes adopted in this paper. Note that in the chosen characteristic units the parameters in these two schemes are related by equation (\ref{betalambda}), i.e. the only parameters $\lambda$ and $\beta$ left in the dimensionless equations are inverse to each other. The immediate problem is that, when $\lambda$ and $\beta$ are assigned relatively large values during the numerical procedure, the corresponding trapping or interaction energy terms will be very large in units of the chosen characteristic wave function $\psi_0$ while the step sizes of spatial and temporal discretization in numerical computation are defined in terms of units $x_0$ and $t_0$, making it difficult to obtain an accurate ground state and thus reliable small-amplitude excitations. Thus we choose $\lambda$-scheme to carry out the calculation for very weak traps as the case of self-bound droplet $\lambda=0$ may serve as the benchmark, while it is more suitable to adopt the $\beta$-scheme for very strong confinement and the trapped ideal gas limit $\omega_B=2\lambda$ is a good reference. In this context, we have checked the breathing modes calculated by these two schemes over the intermediate regions of the parameter $\{\lambda, \beta \}\sim\{0.1,10 \}$ from a numerical perspective. The results of the two eGPEs are nearly identical when either $\lambda$ or $\beta$ is not greater than 10.


\section{Conclusion}\label{S6}

In conclusion, our study provides a comprehensive analysis of the stationary and excitation properties of a one-dimensional quantum droplet by emphasizing the role of a harmonic trapping potential. To address different energy scales, we have introduced two different dimensionless time-dependent Gross-Pitaevskii equations (GPE) with a parametric duality that allows us to study the system under weak and strong trapping potentials, respectively. By means of these equations, we have explored the ground-state properties such as the density, mean-square radius and chemical potential, etc. We find the fragile feature of the flap-top shape in the density distribution with respect to the external potential, and a minimum size in terms of the mean-square radius with respect to the normalization.
	
Furthermore, after applying the Bogoliubov theory or adding a small fluctuation in the GPE, we have carefully discussed the low-lying elementary collective modes, especially the breathing mode in the excitation energy spectrum. By varying the external potential strength, we have shown an intriguing non-monotonic behavior in the breathing-mode frequency which can effectively recovers the results of a 1D uniform quantum droplet, an ideal gas and a conventional trapped BEC in the specific parameter limits. In addition, the breathing-mode frequency is depicted as a function of the particle number and exhibits a maximum value at a critical position, which is closely associated with the minimum in the mean-square radius. The critical particle number in both breathing-mode frequency and the radius is further shown as an exponentially decaying function of the external potential strength. The predicted stationary and excitation properties are directly accessible with current techniques in ultracold quantum gases experiments \cite{Science2018,PRL.120.135301,PRL.120.235301,PRR.1.033155}.

\begin{acknowledgments}
We acknowledge support from the NSF of China (Grant Nos. 12074340 and 12204413) and the Science Foundation of Zhejiang Sci-Tech University (Grant Nos. 20062098-Y and 21062339-Y). 
\end{acknowledgments}

\bibliography{ref}
\end{document}